\documentclass[aps,prl,twocolumn,groupedaddress,showpacs,floatfix]{revtex4}

\usepackage{graphicx}
\usepackage{natbib}

\begin{document}

\title{Forbidden Transitions in a Magneto-Optical Trap}

\author{M. Bhattacharya}
\author{C. Haimberger}
\author{N. P. Bigelow}
\affiliation{Department of Physics and Astronomy and Laboratory
for Laser Energetics, University of Rochester, Rochester, New York
14627}

\date{\today}

\begin{abstract}
We report the first observation of a non-dipole transition in an
ultra-cold atomic vapor. We excite the 3P-4P electric quadrupole
(E2) transition in $^{23}$Na confined in a Magneto-Optical
Trap(MOT), and demonstrate its application to high-resolution
spectroscopy by making the first measurement of the hyperfine
structure of the 4P$_{1/2}$ level and extracting the magnetic
dipole constant A $=$ 30.6 $\pm$ 0.1 MHz. We use cw OODR
(Optical-Optical Double Resonance) accompanied by photoinization
to probe the transition.
\end{abstract}


\maketitle

One of the frontiers in atomic physics is the detection of the
signature of a transition that is classified as ``forbidden."
Forbidden-transition spectroscopy now plays a central role in
tests of fundamental symmetries of Nature. A significant example
is the study of parity non-conserving (PNC) interactions which
provides one of the best measurements of electro-weak symmetry
breaking \cite{wie}. More generally, forbidden transitions have
been experimentally studied \cite{fts} using a variety of
techniques (such as electron impact and laser excitation), in a
range of contexts (from nebular spectra to cold-ion frequency
standards ), and in a number of atoms, ions, and molecules. In the
case of alkali atoms, the first observation of a forbidden
transition dates back to the early days of quantum mechanics
\cite{datta}. More recently, non-dipole effects have been explored
in photoionization \cite{qdi}, n-wave mixing (NWM) \cite{K6nwm}
and collision-induced absorption \cite{uedafukuda}. In the
particular case of sodium, the 3S-3D E2 transition moment has been
measured using NWM \cite{shen} and the 3P-(5P, 4F) transitions
were observed in OODR \cite{lambro}, both via pulsed laser
excitation.

Another frontier in atomic physics is laser cooling and trapping
of atoms. The availability of ensembles of cold atoms has made
accessible entirely new regimes of atomic behavior ranging from
atom-optical effects \cite{atomoptics} to the formation of a
Bose-Einstein condensate\cite{BEC}. Moreover, a cold vapor is a
nearly ideal enabler for precision measurement applications such
as metrology \cite{clock} and high-resolution spectroscopy
\cite{jimna2}. In this Letter, we describe the first experimental
observation of a forbidden atomic transition in a laser-cooled
vapor, confined in a MOT. Our experiment combines the Doppler-free
nature of the MOT, and its specific optical pumping properties,
with the high resolution afforded by cw lasers and the near-unity
efficiency of ion detection. To illustrate the power of this
approach, we demonstrate the electric quadrupolar nature of the
3P-4P transition and use it to analyze the hyperfine structure of
the 4P$_{1/2}$ level. Measurements of hyperfine splittings are of
interest because they are sensitive to electronic correlations and
relativistic effects, providing a benchmark for testing the
accuracy of many-body atomic structure calculations
\cite{arimondo}.

\begin{figure}
\includegraphics{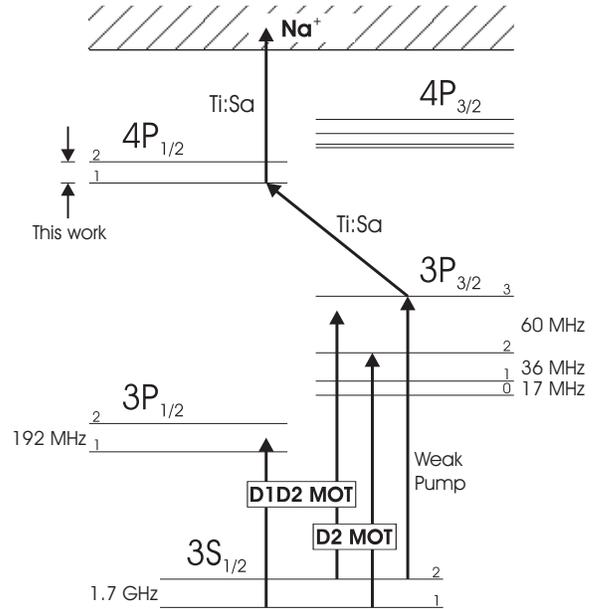}
\caption{\label{ELD}Energy level diagram of sodium showing
trapping configurations and OODR hyperfine splitting measurement.}
\end{figure}

Our initial observations were made with a standard MOT
\cite{jimna2} (Fig.\ref{ELD}) operating on the D2 transition. We
refer to this as the D2MOT.  Atoms held in the trap were probed
with light tunable around 750 nm generated using an Ar$^+$- pumped
Ti:Sa ring laser. While the trapping light excited atoms on the D2
(3S$_{1/2}\rightarrow$3P$_{3/2}$) transition, the probe laser
excited the 3P-4P E2 transition. Atoms populating the resulting
doubly excited state were then ionized to S and D states in the
continuum by either a trap or a probe photon.  The resulting
Na$^+$ ions were detected using a channel electron multiplier
(CEM).

Our experiments were typically carried out using probe laser
powers $\sim$ 100 mW  focused down to a $\sim$ 100 $\mu$m diameter
spot yielding an excitation intensity of $\sim 10^6 {\rm mW/cm}^2$
(the MOT diameter is $\sim$ 300 $\mu$m). The intensity I$_{Sat}$
for saturating the E2 transition can be estimated in the
semi-classical approximation by equating the 3P-4P quadrupole Rabi
frequency ( $\Omega_{Q} \sim QkE_{o}/\hbar$, where $Q$ is the
quadrupole moment of the transition, and k and E$_{o}$ are the
wavevector and electric field amplitude of the 750 nm light ) with
the sum of the linewidths of the 3P and 4P levels ( $\gamma_{D}$
), which are set by dipole (E1) emission to the 3S ground state.
We find \begin{equation} I_{Sat} = \frac{E_{o}^{2}}{2\eta_{o}}
\sim \frac{1}{2\eta_{o}} {\left(
\frac{\hbar\gamma_{D}}{Qk}\right)}^2
\end{equation}
where $\eta_{o}=377$ $\Omega$ is the impedance of free space. With
k=$2\pi$/(750 nm), $\gamma_{D} \sim$ 15 MHz \cite{3P4Pandnote} and
Q $\sim 30$ in atomic units \cite{Qvalue} we arrive at I$_{Sat}
\sim$ 10$^{6}$ mW/cm$^2$, indicating that the E2 transition is
adequately saturated in our experiment.  We were actually able to
observe clearly resolved spectra with good signal-to-noise ratios
down to $\sim$ 20 mW of Ti:Sa power.  It is important to note that
saturation intensities comparable to those quoted here will also
apply to transition moments to higher P states in Na \cite{Qvalue}
as well as in other alkalis \cite{Qsforall}.  This implies that
our excitation techniques can be readily generalized to other
levels and other species. Moreover, the low excitation power
required means that similar E2 studies can be carried out even
using inexpensive, low-power cw-diode lasers familiar to most
cooling and trapping laboratories.

Our subsequent experiments were carried out using a MOT (Fig.
\ref{ELD}), operating on both the D1 and D2 transitions. We refer
to this trap as the D1D2MOT.  In the D1D2MOT, the trapping light
was produced by one dye laser detuned 16 MHz below the
3S$_{1/2}$(F=2)$\rightarrow$3P$_{3/2}$(F$'$=3) D2 transition and
locked using saturation spectroscopy. The repumping light was
produced by a second dye laser that was intense enough to be
resonant with both the 192 MHz split
3S$_{1/2}$(F=1)$\rightarrow$3P$_{1/2}$(F$'$=1, 2) D1 transitions
via power-broadening.  The power broadening eliminated the need to
stabilize the repumping laser against long-term drift.
Anti-Helmholtz coils produced a 20 G/cm magnetic field gradient
and the background pressure was $\sim 10^{-9}$ Torr. From
fluorescence measurements we estimated the number of atoms in the
trap to be $\sim 10^{6}$ at densities $\sim 10^{9}$ cm$^{-3}$.

We find that the D1D2MOT is particularly useful for our
experiments for several reasons. First, in the D2MOT, the presence
of the repumping light tuned to the 3P$_{3/2}$ excited state
manifold complicates the spectrum - the ion production channels
involve two intense laser fields interacting with four closely
spaced hyperfine levels.  By moving the repumping frequency to the
3P$_{1/2}$ manifold we isolate the
3P$_{3/2}$(F=3)$\rightarrow$4P$_{1/2}$(F$'$=1, 2) spectrum and
remove any effects associated with the repumper intensity. Second,
in order to control power-broadening in our spectrum and also to
acquire data at detunings both above and below the 3P$_{3/2}$(F=3)
level, we switched the trapping light off, and used a weak tunable
pump to excite the 3S-3P transition.  However, to prevent optical
pumping into the 3S$_{1/2}$(F=1) state from interfering with our
measurement it was necessary to leave the repumper on, a
configuration made possible by our two-laser scheme. Finally, the
D1D2MOT naturally provides a steady-state population in the
3P$_{1/2}$ state, allowing us to investigate all possible
transitions between the 3P$_{J}$ and 4P$_{J'}$ fine structure
levels, as discussed below.

Having observed the peaks in the ion spectrum associated with the
hyperfine structure of the $4P_{1/2}$ level, we carried out
several studies to confirm that excitation was due to an E2
process. To begin with, we noted that (1) the absence of a
J=1/2$\rightarrow$J$'$=1/2 peak in the
3P$_{J}\rightarrow$4P$_{J'}$ spectrum is a direct consequence of
the $\Delta $J selection rules assuming an E2 process in an alkali
atom \cite{sobelman} and (2) the intensity scale agreement
described in Eq.(1) above depends crucially on E2 excitation. We
then eliminated the only two possible processes that could lead to
excitation in absence of an E2 process: Stark-mixing and
collisional coupling. Consider first Stark-mixing.  In our setup,
the only source of an electric field large enough ($\sim$kV/cm) to
produce Stark-mixing is the CEM and associated ion collection
optics.  We switched off the CEM and still observed depletion of
the MOT fluorescence at resonant Ti:Sa frequencies. While the
presence of Stark-mixing implies a quadratic dependence of the
ion-count on the electric field, no change was observed as the
ion-detection-optic fields were varied by more than a factor of
four.  To address the possibility of collisional effects, we
repeated our measurements using the D2MOT in two distinct
configurations that produce atom clouds differing in density by an
order of magnitude. The first configuration, in which the trapping
light was tuned below the 3P$_{3/2}$(F=3) level, produced higher
densities than the second configuration in which the trapping
light was tuned below the 3P$_{3/2}$(F=2) level. In the presence
of collisional excitation, which scales as density squared, the
ion count rate is expected to change by approximately two orders
of magnitude\cite{molproc}. Instead we observed an ion count rate
that increased only as the number of atoms in the interaction
volume defined by the probe laser, or specifically, with a linear
dependence on atomic density.

The quantitative measurement of the 4P$_{1/2}$ hyperfine structure
was made in the following manner (Fig. \ref{ELD}).  The D1D2MOT
trapping laser was passed through an acousto-optic modulator
(AOM$_{trap}$) and the first order beam provided the trapping
light.  Using the modulator, the trapping light was switched on
and off every 100$\mu$s with a 50\% duty cycle, which allowed us
to maintain a trapped atom number of $\sim 10^{5}$. A weak ($\sim$
1 $\mu$W/cm$^{2}$) pump was introduced into the trap by selecting
a small portion of light from the same laser before it passed
through AOM$_{trap}$ and sending it through a separate AOM whose
first order output was tunable from 16MHz below the
3P$_{3/2}$(F=3) level to 64MHz above. This pump field was not
switched, but was instead left on continuously. Further, the pump
beam was retro-reflected to prevent the MOT from being depleted
due to mechanical pushing effects, particularly when the pump
light was tuned near atomic resonance. In all experiments,
repumping light was applied continuously.

\begin{figure}
\includegraphics{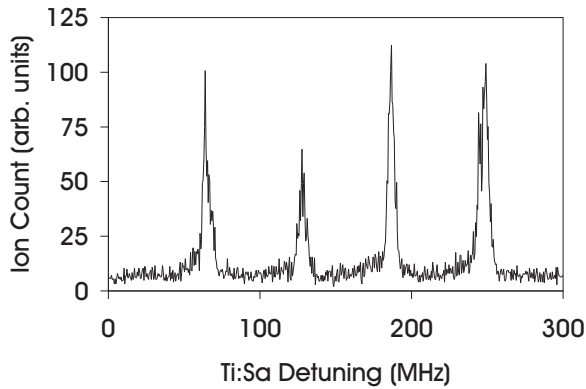}
\caption{\label{SCAN}Typical single scan ion spectrum of
4P$_{1/2}$ hyperfine structure as a function of Ti:Sa frequency,
acquired using the D1D2 MOT.  The first two peaks are due to the
first order output of AOM$_{Ti:Sa}$ upshifted by $\sim$120 MHz
from the zeroth order, which is responsible for the last two
peaks.  The origin of the abscissa is arbitrary.}
\end{figure}

Probe light from the Ti:Sa was also sent through a modulator
(AOM$_{Ti:Sa}$) aligned such that the zeroth and first order beams
had approximately equal power.
 The first order was shifted up in frequency relative to the zeroth order
by $120.0045\pm0.0003$ MHz as measured on a precise frequency
counter.  Both orders were then separately focused on the atoms in
the MOT and each beam contributed a set of peaks to our spectrum
as the Ti:Sa frequency was varied (Fig. \ref{SCAN}).  The offset
supplied by AOM$_{Ti:Sa}$ allowed us to calibrate our measurement
and to check the slope and linearity of the Ti:Sa frequency sweep.
We found the sweep was linear but the nominal value of the slope
required a $3\%$ correction. This procedure eliminated any need to
externally calibrate the Ti:Sa laser frequency.

Using this set-up, we obtained ionization spectra for different
detunings of the pump.  Each spectrum was fit to the sum of four
Lorentzians and a value of the hyperfine splitting was extracted
from each fit. Forty-eight measurements were fit assuming a normal
distribution. We thereby determined a mean value and standard
error for the magnetic dipole constant A(4P$_{1/2}$)= $30.6\pm0.1$
MHz, which is half of the hyperfine splitting.  The theoretical
prediction is 30.7 MHz \cite{safro}, in excellent agreement with
our data.

\begin{figure}
\includegraphics{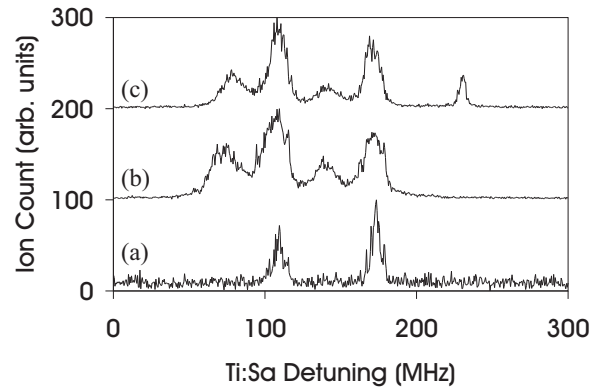}
\caption{\label{AT}Autler-Townes spectrum of 3P$_{3/2}$ manifold
probed from 4P$_{1/2}$(F=1, 2) levels acquired with (a) a
blue-detuned weak pump while the trapping light of the D1D2 MOT is
off, (b) the D1D2 MOT , (c) the D2 MOT. The origin of the abscissa
is arbitrary. }
\end{figure}

When we acquire spectra with the trapping light left on, we
observe Autler-Townes doublets which exhibit dependence on the
Rabi frequency and detuning of the light.  In the case of the
D1D2MOT (Fig. \ref{AT}b) we can identify these peaks as arising
from the interaction of the intense trapping light with the
3P$_{3/2}$(F=3)level. In the case of the D2MOT (Fig. \ref{AT}c)the
same interaction gives rise to all the peaks except for the
rightmost which is due to the repumping light and which
corresponds to the 3P$_{3/2}$(F=2)$\rightarrow$4P$_{1/2}$(F$'$= 2)
transition.  This type of spectra reveals both the effective Rabi
frequency seen by the atoms and the population distribution in the
excited-state of the 3P$_{3/2}$ trapping manifold.  The ability to
detect these populations is of interest, for example, in
elucidating the role of atomic hyperfine structure in
excited-state cold collisions in MOTs \cite{pritchard}. In fact,
E2 transitions are especially well-suited for this type of
measurement for two reasons. First, the $\Delta$F=0, $\pm1, \pm2$
selection rules allow for any single hyperfine level in any higher
P$_{1/2}$ state to probe the entire trapping manifold. More
generally, the weakness of E2 transitions makes them narrower
probes than those that rely on dipole transitions from other
atomic states. It is also interesting to note that, in the context
of a MOT where exchange of momentum with light is important, the
E2 transition is the lowest electric multipolar process in which
exchange of orbital angular momentum between light beams and
internal (as opposed to center-of-mass) atomic motion can be
observed \cite{lh}.

Having demonstrated that E2 excitation of an atomic transition can
be observed using standard cw lasers in combination with a simple
robust laser cooling apparatus, we now turn to extensions and
limitations of our work. A natural first step is to consider the
linewidths extracted from spectra such as in (Fig. \ref{SCAN}). We
observe that the magnitudes of these linewidths are $\sim$10 MHz.
Although the values confirm that the Na atoms are at or below the
Doppler temperature ($\sim 240 \mu$K) the observed linewidth is
convolved with the width of the intermediate 3P$_{3/2}$(F=3)
state, which is known to be $\sim$10 MHz. For more accurate
linewidth measurements, a detuning of the pump field further to
the blue of the 3P$_{3/2}$(F=3) level can be used because in this
limit the spectrum yields the natural linewidth of the 4P state
being ionized. An alternative approach would be to detect the 330
nm fluorescence from the dipolar (E1) decay of the 4P level to the
3S ground state.

If we now consider the peak amplitudes in our spectra we find our
experiment provides a natural way to measure the strength of the
E2 interaction. From the relative heights of the 4P$_{1/2}$ peaks
we estimated the ratio of the quadrupole transition matrix
elements for the F= 1 and 2 states to be 1. Similarly, we compared
the ion yield from the
3S$\rightarrow$3P$\rightarrow$4P$\rightarrow$continuum transition
to that from the dipole-allowed process ( with the probe laser at
820nm ) 3S$\rightarrow$3P$\rightarrow$3D$\rightarrow$continuum.
The ratio of ion counts from the two processes essentially yields
the value of the E2 transition matrix element in terms of the
known 3P$\rightarrow$3D matrix element \cite{lambro}.Our
experimental values agree with the theoretical value used to
estimate Eq.(1) within a factor of 5 uncertainty. Better control
over laser parameters can provide amplitude data of quality high
enough to be compared to theory, and to be of use, for example, in
PNC experiments. We point out that other methods for measuring
forbidden transition matrix elements, such as NWM \cite{shen} and
polarization spectroscopy \cite{havey} have so far been
implemented using pulsed lasers. The application of cw techniques
demonstrated by us would make these methods more accurate and
increase their resolution enabling them to account for atomic
hyperfine structure.

The observation of a forbidden process in a physical system points
to a lowering of symmetry, which in turn implies the presence of
new physics. For example, our observations open up the possibility
of using a MOT to observe second order ($\chi^{(2)}$) nonlinear
processes which are not allowed in the dipole approximation in
centrosymmetric media such as alkali vapors. Also interesting
would be the observation of radiative behavior of cold atoms close
to dielectric surfaces, a situation in which the intensity of
quadrupole transitions has been predicted to become comparable to
that of dipole transitions \cite{d}.The achievement of single-atom
MOTs \cite{s} implies the possibility of PNC measurements via
E1-E2 interference as proposed for Ba$^{+}$ \cite{Ba}.

In conclusion, we report the first observation of a forbidden
transition in a cold atomic vapor formed in a MOT. We use the
resulting excitation process in combination with OODR ionization
spectroscopy to measure the magnetic dipole constant of the
4P$_{1/2}$ level of sodium and find  A $=$ 30.6 $\pm$0.1 MHz. In
addition to demonstrating the application to high-resolution
atomic spectroscopy, we describe how our technique can used to
perform multiphoton ionization and Autler-Townes spectroscopy and
to explore the role of hyperfine structure in alkali MOTs.

\begin{acknowledgments}
We would like to thank Prof. C. Stroud and W. Bittle for
equipment loans.  M.B. would like to thank Prof. J. Muenter for
stimulating discussions. This work was supported by the National
Science Foundation, the Office of Naval Research, the Army
Research Office and the University of Rochester.

\end{acknowledgments}

\bibliography{HFS}

\begin{thebibliography}{25}
\bibitem[{mol()}]{molproc}
\bibinfo{note}{We observe the ion counts scale in this manner for the two MOTs
  when we produce highly excited Na$_{2}$[12].}
\end{thebibliography}

\end{document}